\documentclass[12pt,prd,superscriptaddress,nofootinbib]{revtex4}
\usepackage[english]{babel}
\usepackage{graphicx}
\usepackage{mathtext}
\usepackage{indentfirst}
\usepackage{epsfig,amsmath,amsfonts}

\def\ee{\end{equation}}
\def\ba{\begin{eqnarray}}
\def\ea{\end{eqnarray}}

\def\bq{\begin{quote}}
\def\eq{\end{quote}}

 at 10truept

\newcommand{\beq}{\begin{equation}}
\newcommand{\eeq}{\end{equation}}
\newcommand{\beqa}{\begin{eqnarray}}
\newcommand{\eeqa}{\end{eqnarray}}
\newcommand{\bea}{\begin{eqnarray}}
\newcommand{\eea}{\end{eqnarray}}

\def\ltap{\ \raise.3ex\hbox{$<$\kern-.75em\lower1ex\hbox{$\sim$}}\ }
\def\gtap{\ \raise.3ex\hbox{$>$\kern-.75em\lower1ex\hbox{$\sim$}}\ }
\def\gl{\ \raise.5ex\hbox{$>$}\kern-.8em\lower.5ex\hbox{$<$}\ }
\def\roughly#1{\raise.3ex\hbox{$#1$\kern-.75em\lower1ex\hbox{$\sim$}}}

\newcommand\bqa {\begin{eqnarray}}
\newcommand\eqa {\end{eqnarray}}
\newcommand{\bec}{\begin{cases}}
\newcommand{\eec}{\end{cases}}
\newcommand{\bei}{\begin{itemize}}
\newcommand{\eei}{\end{itemize}}
\newcommand{\bee}{\begin{enumerate}}
\newcommand{\eee}{\end{enumerate}}

\newcommand\pr {\partial}
\newcommand{\fr}{\frac}

\newcommand\nn {\nonumber}

\newcommand{\bear}{\begin{array}}
\newcommand{\enar}{\end{array}}

\newcommand{\tbf}{\textbf}

\begin{document}

\def\I{{\rm i}}

\def\h{\hbar}

\def\t{\theta}
\def\T{\Theta}
\def\w{\omega}
\def\ov{\overline}
\def\a{\alpha}
\def\b{\beta}
\def\g{\gamma}
\def\s{\sigma}
\def\l{\lambda}
\def\wt{\widetilde}
\def\t{\tilde}

\hfill ITEP--TH--5/12

\hfill AEI--2012--011

\vspace{5mm}

\title{\Large Solution of the Dyson--Schwinger equation \\ on de Sitter background in IR limit}
\author{\bf E.\ T.\ Akhmedov}
\affiliation{B.\ Cheremushkinskaya, 25, ITEP, 117218, Moscow, Russia}
\affiliation{Moscow Institute of Physics and Technology, Dolgoprudny, Russia}
\affiliation{Max-Planck-Institut f\"ur Gravitationsphysik (Albert-Einstein-Institut),
Am M\"uhlenberg 1, 14476 Golm, Germany}
\author{\bf Ph.\ Burda}
\affiliation{B.\ Cheremushkinskaya, 25, ITEP, 117218, Moscow, Russia}
\affiliation{Max-Planck-Institut f\"ur Gravitationsphysik (Albert-Einstein-Institut),
Am M\"uhlenberg 1, 14476 Golm, Germany}

\maketitle

\begin{center}{\bf Abstract}\end{center}
We propose an ansatz which solves the Dyson--Schwinger equation
for the real scalar fields in Poincare patch of de Sitter space in the IR limit. The Dyson--Schwinger equation for this ansatz reduces to the kinetic equation, if one considers scalar fields from the principal series. Solving the latter equation we show that under the adiabatic switching on and then off the coupling constant the Bunch--Davies vacuum relaxes in the future infinity to the state with the flat Gibbons--Hawking density of out--Jost harmonics on top of the corresponding de Sitter invariant out--vacuum.

\vspace{5mm}

\section{Introduction}

The goal of the present paper is to understand the impact of large IR loop corrections on the vacuum states in field theory on Poincare patch (PP) of de Sitter (dS) space. In \cite{Krotov:2010ma} one--loop correction to the scalar field Wightman function was calculated in PP over the Bunch--Davies (BD) state \cite{Bunch:1978yq}. The calculation was done in the non--stationary (in--in or Schwinger--Keldysh) diagrammatic technick. There are large IR contributions in the one--loop correction even for the very massive fields.

They reveal themselves through the particle creation --- via the vacuum averages $\langle a^+ \, a\rangle$ and $\langle a \, a\rangle$, where $a$ and $a^+$ are annihilation and creation operators.
E.g. for the real massive scalar field theory with the $\lambda \phi^3$ self--interaction one obtains that $\langle a^+_p \, a_p\rangle \propto \lambda^2 \, \log(p\eta)$ and $\langle a_p \, a_{-p}\rangle \propto \lambda^2 \, \log(p\eta)$ as the conformal time approaches the future infinity, $\eta\to 0$. Here $p$ is the modulus of the spatial co--moving momentum.

Similar IR contributions do appear in other field theories in PP independently of the spin of the fields and self--interaction potentials, as long as they do not respect conformal invariance \cite{Woodard}, \cite{Dolgov:1994cq}, \cite{Antoniadis:2006wq}, \cite{Xue:2012wi}, \cite{Giddings:2010ui}.

In \cite{Akhmedov:2011pj} the observations of \cite{Krotov:2010ma} were generalized to the other dS invariant states (so called $\alpha$--vacua \cite{Mottola:1984ar},\cite{Allen:1985ux}) and to the states containing finite densities of particles. Furthermore, in \cite{Akhmedov:2011pj} kinetic equation was derived. Its solution sums up the leading IR contributions in all loops. One of the goals of the present paper is to show explicitly the latter statement, i.e. to derive that kinetic equation directly from the Dyson--Schwinger (DS) equation of the non--stationary diagrammatic technick. In the situation when, due to the large IR effects, in--out S--matrix approach is not appropriate, the description of the physics via the quantum kinetic (Dyson--Schwinger) equation is more suitable, because the latter equation describes the time--evolution of the state occupation numbers.

The situation with the kinetic theory in dS space demands some clarifications.
Tree--level Wightman function for any $\alpha$--vacuum respects whole dS isometry group \cite{Mottola:1984ar},\cite{Allen:1985ux} even if one restricts field theory to the PP, which covers only half of dS. But there are generators of dS isometry group which deform PP. As the result,
this symmetry is naively broken in the vertexes of the loop integrals to a subgroup, respecting only PP.

One can prove\footnote{We would like to thank A.Polyakov for telling us the idea of this proof. Some elements of the proof can be found in \cite{Polyakov:2007mm},\cite{Polyakov:2009nq},\cite{Polyakovtalk}. See as well the discussion in the Appendix.}, however, that for the BD state the variation of the loop contributions, under those isometry transformations which deform PP, does vanish. Hence, exact Wightman function over the BD state depends only on the dS invariant distance between its two arguments. But for the other $\alpha$--vacua the isometry is broken in loop integrals down to the subgroup in question.

In curved space--times (or in flat space curvilinear coordinates) various coordinate systems frequently cover only their parts. Hence, to do the calculations in such coordinates one has to specify suitable conditions at the boundaries of the corresponding patches. Obviously large IR effects are sensitive to the boundary conditions. Hence, if one does not perform a careful study of the matching between the boundary conditions, he obtains different physical results by doing calculations in different coordinate systems.

In particular it happens that loop contributions in the {\it global} dS space are not just large, but they are explicitly IR divergent even for the massive fields \cite{Akhmedov:2008pu}, \cite{Krotov:2010ma}. In this respect dS space is similar to the QED in strong background electric fields \cite{Akhmedov:2009vh}. The presence of such divergences shows that the moment when the interactions or background field are switched on can not be taken to the past infinity \cite{Krotov:2010ma}. This puts an obstruction for the dS isometry invariance of the correlation functions in global dS and favors the conclusion that cosmological constant should be secularly screened by large IR effects. At least with the appropriate choice of the boundary conditions, i.e. with those boundary conditions which do {\it not} put dS space on the ``life support'' \cite{Krotov:2010ma}.

The fact that dS isometry is respected in the loops over the BD state is a good sign that cosmological constant can not be secularly screened in PP, if the initial conditions are just mild excitations over the BD state. But the presence of the large IR effects means that the BD state itself gets modified.

Indeed, the one--loop correction $G^1(Z)$ to the BD Wightman propagator $G^0(Z)$ is $G^1(Z) \propto \lambda^2\, \log(Z) \, G^0(Z)$,  when the hyperbolic distance is taken to infinity, $Z\to \infty$ \cite{Krotov:2010ma}. This is just the Fourier transform of $\lambda^2 \log(p\eta)$ corrections. Thus, the factor $\lambda^2 \, \log(Z)$ can be big and the loop corrections are not suppressed even if $\lambda^2$ is small.

The question is what is the dressed state? We show below that this question is related to the following one: What is the fate of small density perturbations over the BD vacuum in the future infinity?
To address these questions we derive and solve the kinetic equation which describes the dynamics of such density perturbations and, as we have mentioned, by product sums the leading IR contributions.

From the solution we see that if one sets BD state as the initial one at past infinity, where it is the ground state of the time--dependent free Hamiltonian, this state gets modified even if one switches off the coupling constant at future infinity. It will appear that the result of the summation of all loops will contain modifications of the BD propagator, which do not vanish as $\lambda \to 0$ in the future infinity, but which can not be seen in the free, $\lambda=0$, theory. This makes dS space quite different from Minkowski or Anti--de--Sitter spaces \cite{Akhmedov:2012hk}, where adiabatic variations of the self--interactions do not change the true vacuum state.

To avoid confusions at this point let us clarify our statement. For the fixed co--moving momentum $p$ past infinity in the expanding PP, $\eta \to \infty$, corresponds to the UV limit of the physical momentum, $p\eta$. At the same time future infinity, $\eta\to 0$, corresponds to the IR limit of the physical momentum. So if one starts at the past infinity with the BD state the correlation functions have proper Hadamard UV behavior. What we observe, however, is that for the fixed $p$ as the time goes by, $\eta \to 0$, the IR behavior of the correlation functions is changed (without changing their UV properties) and is described by a different state --- flat density of out--Jost harmonics on top of the corresponding vacuum.

The phenomenon we observe is a more complicated version of the following one. Consider simple linear oscillator. In the perfectly linear case the oscillator will remain in an excited state forever, if it was originally in such a state. However, if one will switch on an interaction of the oscillator to an external field and then switch it off, the oscillator will relax to the ground state. That will happen independently of the type of the interaction or on the type of the external field. The crucial difference of the dS system from the simple oscillator one is that in the case of dS system the oscillator frequency changes in time. As the result even if one had started at past infinity with the ground state of the future infinity, the system would deviate form this state at the intermediate times and then relax back into it in the future.

In the second section we propose the dS invariant Kadanoff--Baym equation which may be suitable to sum the dS invariant IR corrections exactly over the BD state. This section just gives the idea what kind of problem has to be solved if one would like to respect dS isometry exactly. However, we find it rather unphysical to address the question of the stability of the system in the circumstances when all the symmetries are respected exactly. We propose to consider slight excitations above the highly symmetric state and to trace where they evolve in the future infinity. For that reason, in the third section we derive the kinetic equation which does not respect dS isometry, but, unlike full DS equation, is suitable for the separation of the IR renormalization form the UV one. The same equation was derived in \cite{Akhmedov:2011pj}. It was shown there that its collision integral is annihilated by the Gibbons--Hawking density of out--Jost states on top of the out--vacuum. The same state annihilates the collision integral of the Kadanoff--Baym equation of the second section up to subleading terms in IR limit.

To make the paper self--contained we present the general discussion of the scalar fields in PP in the Appendix. All the notations, which are not defined in the main text, can be found in the Appendix.

\section{Towards invariant Kadanoff--Baym equation for BD state}

In this paper we are going to study the following field theory:

\bqa
L = \sqrt{|g|}\,\left[\frac{g^{\mu\nu}}{2}\, \pr_\mu \phi \, \pr_\nu\phi + \frac{m^2}{2}\, \phi^2 + \frac{\lambda}{3}\, \phi^3 + \dots\right].
\eqa
Dots here stand for the higher self--interaction terms, which make the theory stable.
The reason why we are going to consider below formulas only due to the unstable cubic part of the potential is just to simplify them. This instability does not affect our conclusions \cite{Akhmedov:2011pj}.

For the BD state the one loop correction to the Wightman function $G_{-+}$ was calculated in \cite{Krotov:2010ma} (see as well \cite{Leblond}). The result for the sum of the tree--level and one--loop contributions in the IR limit, $Z\to\infty$, is as follows:

\bqa\label{polkrot}
G^{0+1}_{-+}(Z) \approx \left[1 - \frac{\lambda^2\,\left(1 - e^{-2\pi\mu}\right)}{4\,\mu} \, \left|\int_0^\infty dx \, x^{\frac{D-3}{2} - i\, \mu} \, h^2(x)\right|^2 \, \log(Z)\right]\, G^0_{-+}(Z).
\eqa
All notations in this formula and in the formulas that follow are given in the Appendix.

For large enough $D$ the theory in question becomes non--renormalizable. But in the IR limit we do not care about UV divergences and renormalizability of the theory in question. We assume that all couplings in all equations below take their physical values, i.e. all UV divergent ($\sim \lambda^2 \, \log \Lambda$) or finite ($\lambda^2$) contributions are absorbed into their renormalization. For the propagators which have proper Hadamar behavior the UV divergences in dS space are the same as in flat one. Because of that we prefer to consider the BD state (or mild density excitations above it) as the initial state of our system. But below we are keeping track only of the leading large IR contributions.

As seen from (\ref{polkrot}), loops are not suppressed in comparison with the tree--level contribution for large enough $Z$. One has to understand what is the result of the summation of the leading IR contributions at all loops. The answer on this question can be obtained from the solution of the Dyson--Schwinger (DS) equation:

\bqa\label{DS}
\hat{G}(Z_{XY}) = \hat{G}^0(Z_{XY}) + \lambda^2 \int [dW]\int [dU] \hat{G}^0(Z_{XW}) \, \hat{\Sigma}(Z_{WU})\, \hat{G}(Z_{UY}),
\eqa
where $\hat{G}(Z)$ is the matrix of the exact propagators, while $\hat{G}^0(Z)$ is the matrix of the tree--level ones. All propagators in (\ref{DS}) are the functions of the invariant distance, because we are quantizing over the BD state.

Having in mind the physical and mathematical origin of the large IR effects \cite{Akhmedov:2011pj} we have simplified the complete system of DS equations in (\ref{DS}). We have assumed that the vertex $\lambda$ does not receive any new large IR contributions on top of those which are caused by the contributions contained in the two--point functions.

Eq. (\ref{DS}) is not suitable for the summation of only large IR contributions $\lambda^2 \log(Z)$, because it does not separate the UV from the IR renormalization. One needs an equation which sums up only the leading IR contributions and does not even see the contributions which are either suppressed by the higher powers of $\lambda$ or even UV divergent ($\sim \lambda^2 \log \Lambda$). The proper equation is the kinetic one of the next section. However, it does not respect dS isometry, while one would like to sum the dS invariant contributions for the BD state.

One possible variant is as follows. We apply the Klein--Gordon operator to both sides of (\ref{DS}) to get rid of its dependence on the initial value of the propagator. This operator, when acting on the function of $Z$, is equivalent to $\Box(g) + m^2 = (Z^2 - 1) \pr_Z^2 + D\,Z\, \pr_Z + m^2$. That converts the DS equation into an integrodifferential equation of the Kadanoff--Baym form. Recalling that the tree--level Wightman functions, $G^0_{+-}$ and $G^0_{-+}$, solve the homogeneous equation, while the Feynman propagators, $G^0_{++}$ and $G^0_{--}$, solve the inhomogeneous one, we obtain the following equation for the Wightman function $G_{-+}(Z_{XY})$:

\bqa\label{kadbay}
\left[Z_{XY}^2 \pr_{Z_{XY}}^2 + D\,Z_{XY}\, \pr_{Z_{XY}} + m^2\right] \, G[Z_{XY}] = \nn \\ = \lambda^2 \, \int [dW] \, G^2[Z_{XW} + i \epsilon] \, G[Z_{WY} + i \epsilon \, sgn(\eta_w - \eta_y)] + \nn \\ + \lambda^2 \, \int [dW] \, G^2[Z_{XW} + i \epsilon \, sgn(\eta_x - \eta_w)] \, G[Z_{WY} - i \epsilon]
\eqa
in the limit $Z_{XY}\to \infty$. However, we do not see that this equation sums up only the leading IR terms and nothing else. Possible way to move further is to apply the ansatz $G(Z) = f(Z)\, G^0(Z)$ for $Z\to\infty$, where $f(Z)$ is slow in comparison with $G^0(Z)$. But instead we are going to find the stationary IR solution of this equation by approaching the problem from a different perspective.

As a side remark let us mention that it was argued in \cite{Marolf:2010zp} that the result of the summation of the IR contributions should be the propagator build with the use of the exact Hartle--Hawking state. The one which is obtained via analytical continuation from the sphere and constructed with the use of the exact Hamiltonian. Obviously such a state depends on the coupling constant $\lambda$.

The exact state in dS should as well depend on $\lambda$, but besides that we encounter a new phenomenon, which can not be grasped through the analytical continuation from the sphere. We are going to show that the IR stationary solution of (\ref{kadbay}), the one which annihilates its RHS up to subleading terms, does not depend on the coupling constant. I.e. even if we start from the BD state (ground state of the free Hamiltonian on the sphere) and then adiabatically switch on interactions and eventually switch them off the theory relaxes to another state independently from the selfinteractions.

In the next section we will propose the result of the IR dressing of the BD propagator, which, however, will not allow us to fix the function $f(Z)$. Because we will be able to find the propagator at the stationary state, which is reached as $Z\to\infty$, but we will not be able to find the route how it approaches the stationarity in a dS invariant way. We will not be able to find the expression for the propagator at finite values of $Z$. We will find the form of its approach to the stationarity only in the circumstances when the dS isometry is broken.

The hint for the expression of the stationary propagator comes from the following observations. First, the dressed propagator should respect dS isometry. Second, it should annihilate the RHS of (\ref{kadbay}) up to the suppressed terms. These subleading terms can be absorbed into the finite ($\sim \lambda^2$) and infinite ($\sim \lambda^2 \log \Lambda$) UV renormalization.

\section{Solution of the Dyson--Schwinger equation in IR limit}

Let us consider small density perturbation over any $\alpha$--vacuum. Then the dS invariance of the propagators is broken even at tree--level, but we still have large IR contributions.
To sum them up one as well has to solve the DS equation, but this time it does not respect dS isometry.

Due to the relation $G^0_{+-} + G^0_{-+} = G^0_{++} + G^0_{--}$ it is convenient to perform the Keldysh rotation \cite{Kamenev}, \cite{LL} to the new basis: $D^K_0(X,Y) = -\frac{i}{2}\, \left[G^0_{+-}(X,Y) + G^0_{-+}(X,Y)\right]$, $D^R_0(X,Y) = \theta(\eta_y-\eta_x) \, \left[G^0_{-+}(X,Y) - G^0_{+-}(X,Y)\right]$ and $D^A_0(X,Y) = \theta(\eta_x-\eta_y) \, \left[G^0_{+-}(X,Y) - G^0_{-+}(X,Y)\right]$.
Here $D^{R,A}$ are retarded and advanced Green functions. They carry information about the quasi--particle spectrum of the theory. At the same time the Keldysh propagator $D^K$ describes the state of the theory. Thus, our main concern below should be the solution of the DS equation for the Keldysh propagator $D^K$.

Due to spatial homogeneity of PP and due to its rapid expansion, which is supposed to fade away any initial inhomogeneity, we find it convenient to perform the Fourier transform of all quantities along the spatial directions: $D^{K,R,A}_p(\eta_1, \eta_2) \equiv \int d^{D-1}x \, e^{i\, \vec{p}\, \vec{x}} D^{K,R,A}(\eta_1, \vec{x}; \eta_2, 0)$. Then the Fourier transformed form of the DS equation for $D^K$ is as follows\footnote{Feynman rules can be found in \cite{vanderMeulen:2007ah}.}:

\bqa\label{DSDK}
D_p^{K}(\eta_1,\eta_2) =  D_{0p}^K(\eta_1,\eta_2) + \nn \\ + \l^2 \int \fr{d^{D-1}\vec{q}}{(2\pi)^{D-1}} \iint_\infty^0 \fr{d\eta_3 d\eta_4}{(\eta_3\eta_4)^D} \, \Biggl[ \Biggr. D_{0p}^R(\eta_1,\eta_3) \, D_q^K(\eta_3,\eta_4) \, D_{p-q}^K(\eta_3,\eta_4) \, D_p^A(\eta_4,\eta_2) + \nn\\ + 2 \, D_{0p}^R(\eta_1,\eta_3) \, D_q^R(\eta_3,\eta_4) \, D_{p-q}^K(\eta_3,\eta_4) \, D_p^K(\eta_4,\eta_2) + 2 \, D_{0p}^K(\eta_1,\eta_3) \, D_q^K(\eta_3,\eta_4) \, D_{p-q}^A(\eta_3,\eta_4)\, D_p^A(\eta_4,\eta_2) - \nn\\ - \fr{1}{4} \, D_{0p}^R(\eta_1,\eta_3) \, D_q^R(\eta_3,\eta_4) \, D_{p-q}^R(\eta_3,\eta_4) \, D_p^A(\eta_4,\eta_2) - \fr{1}{4} \, D_{0p}^R(\eta_1,\eta_3) \, D_q^A(\eta_3,\eta_4) \, D_{p-q}^A(\eta_3,\eta_4) \, D_p^A(\eta_4,\eta_2) \Biggl. \Biggr].
\eqa
Note that we are looking for the kinetic equation whose collision integral is defined at the $\lambda^2$ order. In such an approximation $D_{0p}^K$ can be substituted by $D_p^K$ under the integral on the RHS of (\ref{DSDK}).

We propose the following ansatz to solve (\ref{DSDK}):

\bqa\label{ansatz}
D_p^{K}(\eta_1,\eta_2) = \left(\eta_1\eta_2\right)^{\fr{D-1}{2}}\, d^K(p\eta_1,p\eta_2) , \nn \\
d^K\bigl(p\eta_1,p\eta_2\bigr) =  \fr12 h\bigl(p\eta_1\bigr)\,h^*\bigl(p\eta_2\bigr) \biggl[1+2\,n\bigl(p\eta_{12}\bigr)\biggr] + h\bigl(p\eta_1\bigr)h\bigl(p\eta_2\bigr)\,\kappa\bigl(p\eta_{12}\bigr) + c.c.,
\eqa
where $\eta_{12} = \sqrt{\eta_1 \eta_2}$. As well we use the tree--level retarded and advanced propagators $D_p^{R}(\eta_1,\eta_2) = \theta\left(\eta_2-\eta_1\right) \, \left(\eta_1\eta_2\right)^{\fr{D-1}{2}} \,
d^-\bigl(p\eta_1,p\eta_2\bigr)$, $D_p^{A}(\eta_1,\eta_2) = - \theta\left(\eta_1-\eta_2\right) \, \left(\eta_1\eta_2\right)^{\fr{D-1}{2}}\,d^-\bigl(p\eta_1,p\eta_2\bigr)$, where $d^-\bigl(p\eta_1,p\eta_2\bigr)= 2\,{\rm Im}\left[h(p\eta_1)h^*(p\eta_2)\right]$. In (\ref{ansatz}) $n(p\eta)$ and $\kappa(p\eta)$ are unknown functions to be defined by the equations under derivation.

This ansatz is inspired by the following observations. The retarded and advanced Green functions can be found as classical objects if the spectrum of quasi--particles is known. The ansatz for the Keldysh propagator follows from the interpretation of $n(p\eta)$ as the particle density, $\langle a^+_p a_p \rangle$, and of $\kappa(p\eta)$ as the anomalous quantum average, $\langle a_p \, a_{-p}\rangle$, \cite{Akhmedov:2011pj}. We assume that in the future infinity $n$ and $\kappa$ are independent of the spatial coordinates. Furthermore, due to the symmetry of the PP under simultaneous rescalings of its coordinates, $\eta \to l\, \eta$ and $\vec{x} \to l \, \vec{x}$, we expect that in the future infinity $n$ and $\kappa$ should be functions of the physical momentum, $p\eta$, only: $n_p(\eta) = n(p\eta)$ and $\kappa_p(\eta) = \kappa(p\eta)$.

It is known in condensed matter physics that non--vanishing $\kappa$ signals that one have chosen wrong harmonics to describe the quasi--particle spectrum. As well for constant $\kappa$ one can always set it to zero by performing Bogolyubov transformation which leads to the same ansatz (\ref{ansatz}), but with harmonics corresponding to a different $\alpha$--vacuum and different value of $n$. Because of these observations we do not specify harmonics until the end where we check the IR behavior of $\kappa(p\eta)$ for the various choices of them.

For general values of $\eta_1$ and $\eta_2$ the ansatz (\ref{ansatz}) does not solve the DS equation in question. However, in the limit $p\eta_{1,2}\to 0$ and $\eta_1/\eta_2=const$, one can neglect the difference between $\eta_1$ and $\eta_2$ in the expressions which follow. That can be done if one keeps track only of the leading large IR contributions.

As a result one can substitute the average conformal time $\eta_{12} = \sqrt{\eta_1\eta_2}$, instead of both $\eta_1$ and $\eta_2$ for the limits of integrations over $\eta_3$ and $\eta_4$. Then the ansatz in question reproduces itself under the substitution into the DS equation if $n$ and $\kappa$ obey:

\bqa\label{np}
n(p\eta_{12}) \approx n_p^{(0)} - \l^2 \, \int \fr{d^{D-1}q}{(2\pi)^{D-1}} \, \iint_{\infty}^{\eta_{12}} d\eta_3 \, d\eta_4 \, (\eta_3\eta_4)^{\fr{D-3}{2}}  \times \nn\\
\times \Biggl\{ \Biggr. \Biggl[ d^K\biggl(q\eta_3,q\eta_4\biggr)\, d^K\biggl(|p-q|\eta_3,|p-q|\eta_4\biggr) + \fr14 \, d^-\biggl(q\eta_3,q\eta_4\biggr) \, d^-\biggl(|p-q|\eta_3,|p-q|\eta_4\biggr)
- \Biggr. \nn \\ - \Biggl. d^-\biggl(q\eta_3,q\eta_4\biggr)\, d^K\biggl(|p-q|\eta_3,|p-q|\eta_4\biggr) \, \biggl[1 + 2\,n\bigl(p\eta_{13}\bigr)\biggr] \Biggr] \, h^*\bigl(p\eta_3\bigr)\,h\bigl(p\eta_4\bigr)
+ \nn \\ + 4 \, \theta\left(\eta_4-\eta_3\right) \, d^K\biggl(|p-q|\eta_3,|p-q|\eta_4\biggr) \, {\rm Re} \biggl[d^-\biggl(q\eta_3,q\eta_4\biggr)\,h\bigl(p\eta_3\bigr)\,h\bigl(p\eta_4\bigr)\, \kappa\bigl(p\eta_{42}\bigr) \biggr] \Biggl. \Biggr\}
\eqa
and
\bqa\label{kappa}
\kappa(p\eta_{12}) \approx \kappa_p^{(0)} - \l^2 \, \int \fr{d^{D-1}q}{(2\pi)^{D-1}} \, \iint_{\infty}^{\eta_{12}} d\eta_3 \, d\eta_4 \, (\eta_3\eta_4)^{\fr{D-3}{2}}  \times \nn\\
\times \Biggl\{ \Biggr. \Biggl[ d^K\biggl(q\eta_3,q\eta_4\biggr) \, d^K\biggl(|p-q|\eta_3,|p-q|\eta_4\biggr) + \fr14 \, d^-\biggl(q\eta_3,q\eta_4\biggr) \, d^-\biggl(|p-q|\eta_3,|p-q|\eta_4\biggr)
+ \Biggr. \nn \\ + \Biggl. d^-\biggl(q\eta_3,q\eta_4\biggr) \, d^K\biggl(|p-q|\eta_3,|p-q|\eta_4\biggr)\, \biggl[1 + 2\,n\bigl(p\eta_{13}\bigr)\biggr] \Biggr] \, h^*\bigl(p\eta_3\bigr) \, h^*\bigl(p\eta_4\bigr)
+ \nn \\ + 4 \, \theta\left(\eta_4-\eta_3\right) \, d^K\biggl(|p-q|\eta_3,|p-q|\eta_4\biggr) \, d^-\biggl(q\eta_3,q\eta_4\biggr) \, h^*\bigl(p\eta_3\bigr) \, h\bigl(p\eta_4\bigr) \, \kappa\bigl(p\eta_{42}\bigr) \Biggl. \Biggr\}
\eqa
where $n_p^{(0)}$ and $\kappa_p^{(0)}$ define the initial propagator $D^K_{0p}(\eta_1, \eta_2)$. Their presence is the drawback of the integral form of the equations under consideration, because then the equation itself depends on the initial conditions. The integrodifferential form of these equations is just the system of kinetic equations for $n$ and $\kappa$ together, which was derived using different methods in \cite{Akhmedov:2011pj}.

In the derivation of (\ref{np}) and (\ref{kappa}) we have used the following relations $d^-(p\eta_1,p\eta_2) = - d^-(p\eta_2,p\eta_1) = - \biggl[d^-(p\eta_1,p\eta_2)\biggr]^*$ and $\int d^{D-1}\vec{q} f\left(q,|p-q|\right) = \int d^{D-1}\vec{q} f\left(|p-q|,q\right)$.
As well we assumed that $n(p\eta)$ and $\kappa(p\eta)$ are slow functions in comparison with $h(p\eta)$. Then one can safely change their positions under the $d\eta_3$ and $d\eta_4$ integrals, what we frequently do in the equations below. This is due to the usual separation of scales, which lays in the basis of the kinetic theory \cite{LL}. In our case this approximation is correct only for the fields from the principal series, $m>(D-1)/2$, for which the harmonics $h(p\eta)$ oscillate at the future infinity.

However, the ansatz (\ref{ansatz}) solves (\ref{DSDK}) as well for the scalars from the complementary series, $m\le (D-1)/2$. For them the harmonics do {\it not} oscillate at future infinity. The main problem with the situation when $h(p\eta)$ is as slow as $n(p\eta)$ and $\kappa(p\eta)$ is that then one can not derive the kinetic equation of the usual form. More complicated integrodifferential equations are available whose solution and physical interpretation is not yet known to us.

To simplify (\ref{np}) and (\ref{kappa}) we change the variables as $q\eta_{1,2,3,4} = x_{1,2,3,4}$ and use some approximations \cite{Akhmedov:2011pj} to arrive at:

\bqa\label{np1}
n(p\eta_{12}) \approx n_p^{(0)} + \frac{\l^2\, S_{D-2}}{(2\pi)^{D-1}} \, \int_p^{1/\eta_{12}} \fr{dq}{q} \, \iint_{\infty}^{0} dx_3 \, dx_4 \, (x_3\,x_4)^{\fr{D-3}{2}}  \times \nn\\
\Biggl\{ \Biggr. \Biggl[ \biggl[d^K\left(x_3,x_4\right)\biggr]^2  +
\fr14 \, \biggl[d^-\left(x_3,x_4\right)\biggr]^2
- d^-\left(x_3,x_4\right)\, d^K\left(x_3,x_4\right) \, \left[1 + 2^{\phantom{\frac12}} n\left(\frac{p}{q}x_{13}\right)\right] \Biggr] \times \nn \\ \times h^*\left(\frac{p}{q}x_3\right)\,h\left(\frac{p}{q}x_4\right) + 4 \, \theta\left(x_4-x_3\right) \, d^K\left(x_3,x_4\right) \, {\rm Re} \left[d^-\left(x_3,x_4\right)\,h\left(\frac{p}{q}x_3\right)\,h\left(\frac{p}{q}x_4\right)\, \kappa\left(\frac{p}{q}x_{42}\right) \right] \Biggl. \Biggr\}\nn \\
{\rm and} \quad
\kappa(p\eta_{12}) \approx \kappa_p^{(0)} - \frac{\l^2\, S_{D-2}}{(2\pi)^{D-1}} \, \int_p^{1/\eta_{12}} \fr{dq}{q} \, \iint_{\infty}^0 dx_3 \, dx_4 \, (x_3x_4)^{\fr{D-3}{2}}  \times \nn\\
\Biggl\{ \Biggr. \Biggl[ \biggl[d^K\left(x_3,x_4\right)\biggr]^2 +  \fr14 \, \biggl[d^-\left(x_3,x_4\right)\biggr]^2 + d^-\left(x_3,x_4\right) \, d^K\left(x_3,x_4\right)\, \left[1 + 2^{\phantom{\frac12}} n\left(\frac{p}{q}x_{13}\right)\right] \Biggr] \times \nn \\ \times h^*\left(\frac{p}{q}x_3\right) \, h^*\left(\frac{p}{q}x_4\right)
+ 4 \, \theta\left(x_4-x_3\right) \, d^K\left(x_3,x_4\right) \, d^-\left(x_3,x_4\right) \, h^*\left(\frac{p}{q}x_3\right) \, h\left(\frac{p}{q}x_4\right) \, \kappa\left(\frac{p}{q}x_{42}\right) \Biggl. \Biggr\}.
\eqa
Here $S_{D-2}$ is the volume of the $(D-2)$--dimensional sphere of unit radius and $x_{ij} = \sqrt{x_i\,x_j}$. In (\ref{np1}) we have neglected $p$ in comparison with $q$ inside the integrals to keep only the leading IR terms. See \cite{Akhmedov:2011pj} for more detailed discussion.

Now for the BD state $h(x)\propto {\cal H}^{(1)}_{i\mu}(x)$. Then the $x_{3,4}$ integrals are saturated around $x\sim \mu$, because of the rapid oscillations of the Hankel function at large values of their arguments. Hence, $h(px_{3,4}/q)$ can be expanded around zero, because $p/q\ll 1$ in (\ref{np1}). Then, because ${\cal H}^{(1)}_{i\mu}(x)$ behaves as $C_+ \, x^{i\mu} + C_- \, x^{-i\mu}$, when $x\to 0$, there are interference terms under the $dq/q$ integral which do not depend on $q$. As the result, both $n(p\eta)$ and $\kappa(p\eta)$ behave as $\lambda^2 \, \log(p\eta)$ in the future infinity. Moreover, $\kappa(p\eta)$ is generated even if it was set to zero at the initial stage \cite{Akhmedov:2011pj}. Its presence in the future infinity signals that the backreaction on the BD state (please do not confuse it with the backreaction on the dS geometry) is huge.

One should be a bit more careful with the similar manipulations for the other $\alpha$--vacua, because their harmonics behave as linear combinations of $e^{ip\eta}$ and $e^{-ip\eta}$ at large momenta.
But the careful study reveals the same picture for the most of them \cite{Akhmedov:2011pj}. The explanation comes from the fact that their harmonics $h(x)$ as well behave as linear combinations of $x^{i\mu}$ and $x^{-i\mu}$ in the future infinity.

Only for the out--Jost harmonics, $h(x) \propto J_{i\mu}(x)$, which behave as single waves $x^{i\mu}$, the situation is different. In particular if one puts $\kappa(p\eta)$ to be zero it is not generated back in (\ref{np1}). Or more precisely, contribution to it behaves as $\lambda^2$, i.e. is negligible in comparison with $\lambda^2\,\log(p\eta)$. That is because the integrand of $dq/q$, defining $\kappa$ in (\ref{np1}), contains only $q$--dependent terms and, hence, the corresponding integral is convergent as $p\eta \to 0$. At the same time for the harmonics in question $n(p\eta)$ has contributions of the order of $\lambda^2 \log(p\eta)$.
(The physics for the general $\alpha$--vacua was discussed in grater details e.g. in \cite{Kundu:2011sg}.)

All in all, out--Jost harmonics represent the proper quasi--particle states in the future infinity. Which means that for out--Jost harmonics the ansatz (\ref{ansatz}) with $\kappa(p\eta)=0$ {\it does} reproduce itself after the substitution into DS equation. That is possible if one neglects terms which are suppressed in comparison with powers of $\lambda^2 \log(p\eta)$. This is the argument which favors the interpretation that independently of the initial state at the past infinity of PP the field theory state flows in the future infinity to the out--vacuum with some density of particles on top of it \cite{Akhmedov:2011pj}. To support such a conclusion we are going to show in a moment that for the out--Jost harmonics $\kappa(p\eta)$ indeed flows to zero in the future infinity, even if it was not zero originally.

The kinetic equation is obtained from (\ref{np1}) when $\kappa(p\eta)$ set to zero, via application of the differential operator to its both sides:

\bqa\label{cint}
\frac{d n(x)}{d\log (x)} = - \frac{\lambda^2\, S_{D-2}}{2 (2\,\pi)^{D-1}\, \mu} \, \int^0_{\infty} dx_3 \,x_3^{\frac{D-3}{2}} \, \int_{\infty}^{0} dx_4 \, x_4^{\frac{D-3}{2}} \times \nonumber \\ \times \left\{ {\rm Re} \left[x_3^{-i\mu} \, V(x_3) \,x_4^{i\mu}\, V^*(x_4)\right] \, \left[(1+n(x))\, n(x_3)^{2\phantom{\frac12}} - \,\, n(x) \, (1+n(x_3))^2\right] + \right. \nonumber \\
+ 2\,{\rm Re} \left[x_3^{i\mu} \, W(x_3) \, x_4^{- i\mu} \, W(x_4) \right]\, \left[ n(x_3)\, (1+n(x_3))\,(1+n(x))^{\phantom{\frac12}} - \,\,(1+n(x_3))\, n(x_3)\, n(x) \right] + \nonumber \\
+ \left. {\rm Re} \left[x_3^{i\mu} \, V(x_3) \, x_4^{- i\mu} \, V^*(x_4)\right] \, \left[ (1+n(x_3))^2 \,(1+n(x))^{\phantom{\frac12}} - \,\, n(x_3)^2\, n(x) \right]^{\phantom{2}}\right\}.
\eqa
Here $x=p\eta_{12}$, $V(x) = \left[h^2\left(x\right) - \frac{\pi \, e^{-\pi\mu}}{4\, \sinh(\pi\, \mu)\, |x|} - \dots\right]$ and $W(x) = \left[\left|h\left(x\right)\right|^2- \frac{\pi \, e^{-\pi\mu}}{4\, \sinh(\pi\, \mu)\, \left|x\right|} - \dots\right]$, where $h(x) = \sqrt{\frac{\pi}{\sinh(\pi\mu)}}\, J_{i\mu}(x)$ with $J$ being  the Bessel function. Dots in these expressions stand for a finite number of terms with higher powers of $1/|x|$. The presence of such contributions makes the collision integral well defined after the Taylor expansion of $h(px/q)$ and can be explained by the behavior of the out--Jost harmonics in the limit $x\to\infty$. All this is clarified in \cite{Akhmedov:2011pj}.

This is exactly the kinetic equation which was derived in \cite{Akhmedov:2011pj}. If one have started with a small density perturbation over the BD vacuum, he can expect that $n(p\eta)$ is small in the future infinity. As is explained in \cite{Akhmedov:2011pj} in this case (\ref{cint}) degenerates into a renormalization group type differential equation. The latter one can be solved with the result:

\bqa\label{solution}
n(p\eta) = \frac{\Gamma_2}{\Gamma_1}\left[C \, \left( p\,\eta\right)^{\Gamma} + 1\right], \nonumber \\
\Gamma_1 = \frac{\lambda^2\, S_{D-2}}{(2\pi)^{D-1}\, \mu} \, \left|\int_0^{\infty} dy \,y^{\frac{D-3}{2} - i\, \mu} \, V(y) \right|^2, \nonumber \\ \Gamma_2 = \frac{\lambda^2\, S_{D-2}}{(2\,\pi)^{D-1}\, \mu} \, \left|\int_0^{\infty} dy \,y^{\frac{D-3}{2} + i\,\mu} \, V(y)\right|^2.
\eqa
where $C$ is the integration constant, which depends on the initial conditions.

This solution has stable point $\frac{\Gamma_2}{\Gamma_1} \approx e^{-2\,\pi\mu}\ll 1$ for $\mu\gg 1$, which approximately annihilates the collision integral in (\ref{cint}). The stable point is reached when the production of particles is equilibrated by their decay \cite{Akhmedov:2011pj}. In fact, from the collision integral (\ref{cint}) it should be clear that $\Gamma_1$ defines the decay rate of the scalar particle into two, while $\Gamma_2$ defines the particle production rate. Note that $\log(p\eta)$ is decreasing as we approach the future infinity and $n(p\eta)$ is the density per co--moving volume, which does not dependent on scale $1/\eta$ \cite{Akhmedov:2011pj}.

What is the most interesting fact, from the perspective of the discussion above, is that the stable point in question does not depend on $\lambda$. (Of cause the way the solution (\ref{solution}) approaches the stationarity (its value for non--zero $p\eta$) does depend on $\lambda$.) Furthermore, it is not hard to see now that by product we have shown that the stationary state of the kinetic equation (\ref{cint}) as well annihilates, modulo subleading terms, the RHS (collision integral) of (\ref{kadbay}).

The last thing which we have to check is the behavior of $\kappa(p\eta)$ for the out--Jost harmonics if it was initially non--zero. We as well assume that we have started from its small value in the past infinity and that it flows to the zero in the future. Under these assumptions, if one keeps only the leading terms, the integrodifferential form of the equation for $\kappa(p\eta)$ from (\ref{np1}) degenerates to:

\bqa
\frac{d\kappa(p\eta)}{d\log(p\eta)} = \Gamma_3 \, \kappa(p\eta), \nn \\
\Gamma_3 = \frac{4\, i\, \lambda^2 \, S_{D-2}}{(2\pi)^{D-1}}\, \iint^0_\infty dx_3 \, dx_4 \, x_3^{\frac{D-3}{2} - i\mu}\, x_4^{\frac{D-3}{2} + i\mu}\, \theta(x_4-x_3) \, {\rm Im}\left[V(x_3)V^*(x_4)\right].
\eqa
Here Re$\Gamma_3 = \Gamma_1 - \Gamma_2 \approx \left(1 - e^{-2\pi\mu}\right)\, \Gamma_1 > 0$ and, hence, the solution of this equation, $\kappa(p\eta) \propto \left(p\eta\right)^{\Gamma_3}$, flows to zero in the future infinity. I.e. our assumption is self consistent and (\ref{solution}) is stable under linearized perturbations of $\kappa(p\eta)$.

\section{Conclusions and Acknowledgments}

We have found the result of IR dressing of the BD vacuum in PP of dS. The dressed state is
described by out--Jost harmonics and corresponds to $\kappa(p\eta)=0$ with $n(p\eta) \approx e^{-2\pi\mu}$. Furthermore, the corresponding two--point correlation function depends, in the future infinity, only on the time difference, $\eta_1/\eta_2 = e^{t_2 - t_1}$, rather than on both of the times ($\eta_1$ and $\eta_2$) independently. Which means that the dressed state as well solves the kinetic problem in dS space.

Using the same methods as those which lead to (\ref{cint}) one can derive the kinetic equation in the contracting PP, $ds^2 = dt^2 - e^{-2t}\, d\vec{x}^2 = \frac{1}{\eta^2}\, \left(d\eta^2 - d\vec{x}^2\right)$ where $0\to \eta = e^t \to +\infty$. The solution of the latter equation for low momenta $p$ is \cite{Akhmedov:2011pj}:

\bqa
n(\eta) \sim \frac{1}{A - \bar{\Gamma}\, \log\eta} \sim \frac{1}{\bar{\Gamma}\,\log\frac{\eta_0}{\eta}},
\eqa
and is independent of $p$. It is valid for $\eta < \eta_0 = e^{const/\lambda^2}\gg 1$. Here $A$ is an integration constant, which depends on the initial state and $\bar{\Gamma} \propto \frac{\lambda^2}{m^2} > 0$ for $m \gg (D-1)/2$.

One can see that the distribution in question grows with time, due to the contraction of the space and constant particle production, and moreover has a pole at some finite $\eta_0$. In this case the backreaction on the gravitational background should be strong.
This observation means that in global dS space the situation can be quite different form the one in the expanding PP, at least because global dS contains expanding and contracting PP simultaneously. Then we have two competing processes --- expansion of the space time and explosive particle production \cite{Akhmedov:2011pj}.

We would like to acknowledge discussions with A.Polyakov and I.Burmistrov. We would like to thank MPI, AEI, Golm, Germany for the hospitality during the final stage of the work on this project. The work of AET was partially supported by the grant "Leading Scientific Schools" No. NSh-6260.2010.2, RFBR-11-02-01227-a. The work of PhB was partially supported by the grant RFBR--11--02--01120 by the Dynasty Foundation. This work was done under the support of the grant from the Ministry of Education and Science of the Russian Federation, contract No. 14.740.11.0081.

\section{Appendix}

The $D$--dimensional de Sitter (dS) space is the hyperboloid, $X_\mu^2 \equiv - X_0^2 + X_i^2 = 1$, ($\mu = 0, 1, \dots, D$ and $i=1,\dots,D$) in the $(D+1)$--dimensional Minkowski space $ds^2 = dX_0^2 - dX_i^2$. Throughout this paper we fix the curvature of the hyperboloid to be one. The expanding Poincare patch (PP) of this space is defined by the coordinates:

\bqa \label{induced}
X_0 = \sinh t + \frac{\vec{x}^2}{2}\, e^t, \quad X_D = - \cosh t + \frac{\vec{x}^2}{2}\, e^t \nn \\
X_a = e^t \, x_a, \quad a=1,\dots, D-1
\eqa
and covers only half of dS space, $X_0 - X_D= e^t \geq 0$. The induced metric in these coordinates is $ds^2 = dt^2 - e^{2\,t}\, d\vec{x}^2 = \frac{1}{\eta^2}\, \left(d\eta^2 - d\vec{x}^2\right)$, where $\eta = e^{-t} = 1/(X_0 - X_D)$. The past infinity of the PP corresponds to $t\to -\infty$, i.e. to $\eta = + \infty$. This is the boundary of the PP inside global dS space, $X_0 = X_D$. The future infinity is at $t=+\infty$, i.e. at $\eta = 0$. The dS isometry is just the rotation symmetry group of the ambient Minkowski space, $SO(D,1)$.

To quantize scalar fields in PP one has to specify time--dependent part of the harmonics $g_p(\eta) = \frac{\sqrt{\pi}\, \eta^{\frac{D-1}{2}}}{2} \, h(p\eta)$ inside the harmonic expansion $\phi(\eta, \vec{x}) = \int d^{D-1}p \, \left[a_p \, g_p(\eta) \, e^{-i\, \vec{p}\, \vec{x}} + a^+_p \, g^*_p(\eta) \, e^{i\, \vec{p}\, \vec{x}}\right]$. From the Klein--Gordon equation in PP follows that $h(p\eta)$ has to solve Bessel equation with the index $\mu = \sqrt{m^2 - \left(\frac{D-1}{2}\right)^2}$, where $m$ is the mass of the particle.

In time dependent backgrounds there is no basis of harmonics which can diagonalize the free Hamiltonian once and forever. The choice of the harmonics in the calculations corresponds to the choice of the background state $a_p \, |vac\rangle = 0$, usually referred to as vacuum. Bunch--Davies (BD) vacuum \cite{Bunch:1978yq} corresponds to $h(p\eta) = e^{-\frac{\pi\mu}{2}}\, {\cal H}_{i\mu}^{(1)}(p\eta)$, where ${\cal H}^{(1)}$ is the Hankel function of the first kind. These harmonics behave as $e^{ip\eta}$ at the past infinity and diagonalize the free Hamiltonian only in that part of space--time.

The other so called $\alpha$--vacua can be obtained form the BD one via the corresponding Bogolyubov transformations and correspond to the harmonics which are linear combinations of the Hankel functions of both kinds ${\cal H}^{(1)}$ and ${\cal H}^{(2)}$. See e.g. \cite{Allen:1985ux} for a similar discussion in the global dS coordinates.

Because of the time dependence of the Hamiltonian one has to apply
the Schwinger--Keldysh diagrammatic technic instead of the Feynman one. In this technic every particle is characterized by the matrix of four propagators (see e.g. \cite{Kamenev}, \cite{LL}):

\bqa
G^0_{-+}(X,Y) = i\,\langle \phi(X) \phi(Y) \rangle, \quad  G^0_{+-}(X,Y) = i\,\langle \phi(Y) \phi(X) \rangle, \nn \\ G^0_{++}(X,Y) = \langle T \,\phi(X) \phi(Y) \rangle = \theta(\eta_y - \eta_x) \, G^0_{-+}(X,Y) + \theta(\eta_x - \eta_y)\, G^0_{+-}(X,Y), \nn \\
G^0_{--}(X,Y) = \langle \bar{T} \,\phi(X) \phi(Y) \rangle = \theta(\eta_y - \eta_x) \, G^0_{+-}(X,Y) + \theta(\eta_x - \eta_y)\, G^0_{-+}(X,Y),
\eqa
which obey one relation $G^0_{+-} + G^0_{-+} = G^0_{++} + G^0_{--}$.
Here ($\bar{T}$) $T$ is the (anti--)time ordering. Note that the conformal time in our definition flows in the reverse direction ($\infty \to \eta \to 0$) with respect to the ordinary time $t$.

All these propagators can be written with the use of the
Wightman function $G^0(X,Y) \equiv i\,\langle \phi(X) \phi(Y)\rangle$. The latter solves the Klein--Gordon equation in the metric of PP. This equation is invariant under the full dS isometry although coordinates (\ref{induced}) are restricted only to the half of dS space. Hence, the solution of the Klein--Gordon equation should depend on the invariant distance between its two arguments --- the two points on the hyperboloid, $X_\mu^2 = 1$ and $Y_\mu^2 = 1$. The convenient function of the latter one on the hyperboloid is the so called hyperbolic distance $Z = - X_\mu \, Y^\mu$. As follows from (\ref{induced}) it is equal to $Z = 1 + \frac{(\eta_x - \eta_y)^2 - |\vec{x} - \vec{y}|^2}{2\eta_x \, \eta_y}$ in PP.

The Klein--Gordon operator, when acting on the function of $Z$, rather than on the function of the two points $X$ and $Y$ separately, is equivalent to  $\Box(g) + m^2 = (Z^2 - 1) \pr_Z^2 + D\,Z\, \pr_Z + m^2$ \cite{Allen:1985ux}, \cite{Mottola:1984ar}. After the change of variables to $x=(1+Z)/2$ the Klein--Gordon equation acquires the form of the hypergeometric one. Its solution (away from the singularity) is the following linear combination of the$\phantom{1}_{2}F_1$ hypergeometric functions:

\bqa\label{green}
G^0(Z) = A_1 \, F\left(\frac{D-1}{2} + i\mu, \frac{D-1}{2} - i\mu; \frac{D}{2}; \frac{1+Z}{2}\right) + \nn \\ + A_2 \, F\left(\frac{D-1}{2} + i\mu, \frac{D-1}{2} - i\mu; \frac{D}{2}; \frac{1-Z}{2}\right), \quad
\mu = \sqrt{m^2 - \left(\frac{D-1}{2}\right)^2}.
\eqa
Here $A_{1,2}$ are some constants which depend on the choice of the $\alpha$--vacuum state with respect to which the averaging is done (see e.g. \cite{Allen:1985ux}). For the BD vacuum $A_2=0$.
To take care of the behavior of this function at its poles and to obtain it as the quantum average $\langle \phi(X) \phi(Y)\rangle$ one has to be more careful.

In fact, Green function (\ref{green}) has three singular points in the complex $Z$--plane: $Z=-1,1,\infty$. They correspond to the usual singular points $x\equiv (1+Z)/2 = 0,1,\infty$ of the hypergeometric equation. The singular behavior $G^0(Z)\propto 1/(Z-1)^{D/2 - 1}$ corresponds to the situation when $X$ and $Y$ sit on the same light--cone --- the standard UV singularity of the propagator. Similar singularity of $G^0(Z)$ at $Z=-1$ corresponds to the situation when $X$ sits on the light--cone with the apex at the antipodal point of $Y$. The antipodal point is obtained via the reflection at the origin of the ambient Minkwoski space \cite{Allen:1985ux}. Finally, at the infinity $G^0(Z)$ has the branching point: $\lim_{Z\to\infty} G^0(Z) \propto Z^{-\frac{D-1}{2}} \, \left[C_1 \, Z^{i\,\mu} + C_2\, Z^{- i\, \mu}\right]$ with some constants $C_{1,2}$.

To understand the behavior of $G^0(Z)$ at its poles it is instructive to consider Fourier transform of $G^0(Z)$ along the homogeneous spatial directions:

\bqa\label{Fourier}
\left\langle \phi\left(\eta_x, \vec{p}\right) \phi\left(\eta_y, -\vec{p}\right)\right\rangle \equiv \int d^{D-1}x \, e^{i\, \vec{p}\, \left(\vec{x} - \vec{y}\right)}\, G^0(Z) = \frac{\left(\eta_x\, \eta_y\right)^{\frac{D-3}{2}}}{2}\, h(p\eta_x)\, h^*(p\eta_y),
\eqa
The appearance of different solutions of the Bessel equation in place of $h(p\eta)$ here is in one--to--one correspondence with the concrete values of $A_{1,2}$ in (\ref{green}) \cite{Allen:1985ux}.

Let us consider the BD propagator. Its only singularity inside the complex $Z$--plane is at $Z=1$ and corresponds to the limit $p\to\infty$ in momentum space. In this limit the Hankel functions behave as the plane waves. As it should be high momentum modes are not sensitive to the curvature of the space--time, i.e. they coincide with the flat space harmonics.
For the inverse of the transformation (\ref{Fourier}) to be well defined there should be an appropriate shift as $\eta_x - \eta_y \to \eta_x - \eta_y \pm i\, \epsilon$ in (\ref{Fourier}). The sign of this shift depends on which one among $\eta_x$ and $\eta_y$ is grater. As the result for the BD state \cite{Polyakov:2007mm}:

\bqa
G^0_{++}[Z] = G^0[Z - i\, \epsilon], \quad G^0_{+-}[Z] = G^0[Z - i \, \epsilon \, sgn(\eta_x - \eta_y)], \nn \\ G^0_{--}[Z] = G^0[Z + i\, \epsilon], \quad G^0_{-+}[Z] = G^0[Z + i \epsilon \, sgn(\eta_x - \eta_y)].
\eqa
Here $G^0(Z)$ is analytic on the complex $Z$--plane with the single cut going from $Z=1$ to infinity along the real axis.

The situation for the other $\alpha$--vacua is different because in those situations harmonics are linear combinations of the Hankel functions of the two kinds ${\cal H}^{(1)}$ and ${\cal H}^{(2)}$. The latter behave at large momenta as $e^{-ip\eta}$ instead of $e^{ip\eta}$. As the result for the other $\alpha$--vacua $G^0(Z)$ is defined on the complex $Z$--plane with two cuts connecting $Z=1$ and $Z=-1$, correspondingly, to infinity and going, due to the $i\epsilon$ shifts, in the opposite halfs of the complex $Z$--plane.

Let us say a few words about the one--loop contribution to the propagators due to the $\lambda \, \phi^3$ self--interaction. In the Schwinger--Keldysh diagrams there are two types of the vertices: of the ``$+$'' and ``$-$'' type, correspondingly. In the ``$+$'' (``$-$'') type vertex only ``$+$'' (``$-$'') ends of the propagators can terminate. Correspondingly the one loop correction can be written as:

\bqa\label{oneloop}
\hat{G}^1(Z_{XY}) = \lambda^2 \, \int [dW] \int [dU]\, \hat{G}^0(Z_{XW}) \, \hat{\Sigma}^0(Z_{WU}) \, \hat{G}^0(Z_{UY}) \eqa
where

\bqa
\hat{G}^{0,1}(Z) = \left(
  \begin{array}{cc}
    G^{0,1}_{--}(Z) & G^{0,1}_{-+}(Z) \\
    G^{0,1}_{+-}(Z) & G^{0,1}_{++}(Z) \\
  \end{array}
\right), \quad {\rm and} \quad \hat{\Sigma}^0(Z) = \left(
  \begin{array}{cc}
    \left[G^0_{--}(Z)\right]^2 & \left[G^0_{-+}(Z)\right]^2 \\
    \left[G^0_{+-}(Z)\right]^2 & \left[G^0_{++}(Z)\right]^2 \\
  \end{array}
\right)
\eqa
and  the measure is $[dW] = d^{(D+1)} W \, \delta\left(W_\mu^2 - 1\right)\,\theta \left(W_0 - W_D\right)$, which is equivalent to the measure $\frac{d\eta}{\eta^D}\, d^{D-1}x$ on PP.
This formula for $\hat{G}^1$ is valid for any $\alpha$--vacuum. Note that in (\ref{oneloop}) dS isometry is naively broken by the presence of the Heavyside $\theta$--function in the integration measure, which restricts to the PP.

But let us examine how $\hat{G}^{1}$ does change under those transformations of $SO(D,1)$ which change the argument of the $\theta$--function. (Here we reproduce the arguments of \cite{Polyakovtalk}.) Let us perform an infinitesimal rotation around $X_0$ towards say $X_1$: $X_D \to X_D - \varphi \, X_1$. Taylor expanding the integration measure up to the first order in $\varphi$, we get: $\delta\, \int[dW]\dots = \int d^{(D+1)} W \, \delta\left(W_\mu^2 - 1\right)\,\delta\left(W_0 - W_D\right)\, \varphi\, W_1 \dots = \int d(W_0 + W_D) \, d^{(D-1)}W \, \delta\left(W_\mu^2 - 1\right)\,\varphi\, W_1 \dots$.

Hence, the contribution of one diagram form (\ref{oneloop}) to the variation of say $G^1_{+-}$ over the BD vacuum state is as follows:

\bqa
\delta_{fisrt} G^{1}_{+-}(X,Y) = \nn \\ \lambda^2 \, \varphi\,\int d^{(D+1)} W \, \delta\left(W_\mu^2 - 1\right)\,\delta\left(W_0 - W_D\right)\, W_1 \int [dU]\times \nn \\ \times G\left[Z_{XW} - i\, \epsilon\right] \, G^2\left[Z_{WU} - i \, \epsilon\right] \, G\left[Z_{UY} - i \epsilon \, sgn\left(\frac{1}{U_0 - U_D} - \frac{1}{Y_0-Y_D}\right)\right] + \nn \\
+ \lambda^2 \, \varphi\,\int [dW] \int d^{(D+1)} U \, \delta\left(U_\mu^2 - 1\right)\,\delta\left(U_0 - U_D\right)\, U_1\times \nn \\ \times G\left[Z_{XW} - i\, \epsilon\right] \, G^2\left[Z_{WU} - i \, \epsilon\right] \, G\left[Z_{UY} - i \epsilon \, sgn\left(\frac{1}{U_0 - U_D} - \frac{1}{Y_0-Y_D}\right)\right] = \nn \\
\lambda^2 \, \varphi\,\int d(W_0 + W_D) \, d^{(D-1)} W \, \delta\left(W_\mu^2 - 1\right)\, W_1 \int [dU]\times \nn \\ \times G\left[Z_{XW} - i\, \epsilon\right] \, G^2\left[Z_{WU} - i \, \epsilon\right] \, G\left[Z_{UY} - i \epsilon \, sgn\left(\frac{1}{U_0 - U_D} - \frac{1}{Y_0-Y_D}\right)\right] + \nn \\
+ \lambda^2 \, \varphi\,\int [dW] \int d(U_0+U_D) \, d^{(D-1)} U \, \delta\left(U_\mu^2 - 1\right)\, U_1\times \nn \\ \times G\left[Z_{XW} - i\, \epsilon\right] \, G^2\left[Z_{WU} - i \, \epsilon\right] \, G\left[Z_{UY} - i \epsilon \right]
\eqa
We are going to show now that
both integrals in the last expressions do vanish because the integrands of $d(W_0+W_D)$ and $d(U_0+U_D)$ are analytical functions in the lower complex $(W_0+W_D)$-- and $(U_0+U_D)$--planes, correspondingly.

Let us examine first the situation with the $d(W_0+W_D)$ integral. As we have pointed out above its integrand is analytical in the lower half $Z$-plane, because the cut goes just above the real axis due to the shift by $i\epsilon$ in the arguments of the propagators. At the same time $Z_{XW} = -\frac12\, (X_0-X_D)\, (W_0+W_D) - \frac12\, (X_0+X_D)\, (W_0-W_D) + X_a\, W_a$. But $W_0-W_D=0$, because of the presence of the $\delta(W_0 - W_D)$ in the integration measure for $\delta G^1$ and $X_0 - X_D \ge 0$, because we are in PP. Hence, $G(Z)$ as the function of $W_0 + W_D$ has the same analytical properties as the function of $Z_{XW}$. Furthermore, because propagators have a power like decay as $(W_0+W_D)\to \infty$ one can close the integration contour by the infinite semicircle in the lower half of the complex $(W_0+W_D)$--plane. The integrand is analytical inside the contour. Hence, the integral is zero. Similar arguments work for the $d(U_0 + U_D)$ integral.

Along the same lines one can show that all the contributions to $\delta G^{1}_{+-}$ do vanish. That is true as well for the infinitesimal rotations in the other directions. Hence, in the case of BD state $G^{1}_{+-}(X,Y)$ is invariant under the full dS isometry and is the function of $Z_{XY}$ only. Similarly one can prove the invariance of the one loop contributions to the other propagators in BD vacuum. Furthermore, one can easily extend these arguments to higher loops.

But all this does not work for the other $\alpha$--vacua, because in that case, as we have mentioned, tree--level propagators have another cut going from $Z=-1$ to infinity and it should be shifted to the other half of the complex $Z$--plane. Hence, loop corrections to the propagators in $\alpha$--vacua respect only that subgroup of all dS isometry, which leaves the PP in question invariant.


\begin{thebibliography}{99}

\bibitem{Krotov:2010ma}
  D.~Krotov and A.~M.~Polyakov,
  Nucl.\ Phys.\ B {\bf 849}, 410 (2011)
  [arXiv:1012.2107 [hep-th]].

\bibitem{Bunch:1978yq}
  T.~S.~Bunch and P.~C.~W.~Davies,
  Proc.\ Roy.\ Soc.\ Lond.\  A {\bf 360}, 117 (1978).

\bibitem{Woodard}
  T.~Prokopec, N.~C.~Tsamis and R.~P.~Woodard,
  Annals Phys.\  {\bf 323}, 1324 (2008)
  [arXiv:0707.0847 [gr-qc]];\\
  R.~P.~Woodard,
  J.\ Phys.\ Conf.\ Ser.\  {\bf 68}, 012032 (2007)
  [gr-qc/0608037];\\
  T.~Prokopec, N.~C.~Tsamis and R.~P.~Woodard,
  Class.\ Quant.\ Grav.\  {\bf 24}, 201 (2007)
  [gr-qc/0607094];\\
  S.~-P.~Miao and R.~P.~Woodard,
  Phys.\ Rev.\ D {\bf 74}, 044019 (2006)
  [gr-qc/0602110];\\
  S.~-P.~Miao and R.~P.~Woodard,
  Class.\ Quant.\ Grav.\  {\bf 23}, 1721 (2006)
  [gr-qc/0511140];\\
  N.~C.~Tsamis and R.~P.~Woodard,
  Nucl.\ Phys.\ B {\bf 724}, 295 (2005)
  [gr-qc/0505115];\\
  R.~P.~Woodard,
  Nucl.\ Phys.\ Proc.\ Suppl.\  {\bf 148}, 108 (2005)
  [astro-ph/0502556];\\
  N.~C.~Tsamis and R.~P.~Woodard,
  Nucl.\ Phys.\ B {\bf 474}, 235 (1996)
  [hep-ph/9602315];\\
  N.~C.~Tsamis and R.~P.~Woodard,
  Annals Phys.\  {\bf 253}, 1 (1997)
  [hep-ph/9602316];\\
  N.~C.~Tsamis and R.~P.~Woodard,
  Class.\ Quant.\ Grav.\  {\bf 11}, 2969 (1994).

\bibitem{Dolgov:1994cq}
  A.~D.~Dolgov, M.~B.~Einhorn and V.~I.~Zakharov,
  Phys.\ Rev.\  D {\bf 52}, 717 (1995)
  [arXiv:gr-qc/9403056].

\bibitem{Antoniadis:2006wq}
  I.~Antoniadis, P.~O.~Mazur and E.~Mottola,
  New J.\ Phys.\  {\bf 9}, 11 (2007)
  [arXiv:gr-qc/0612068];\\
  E.~Mottola,
  Phys.\ Rev.\  D {\bf 33}, 1616 (1986);\\
  E.~Mottola,
  Phys.\ Rev.\  D {\bf 33}, 2136 (1986).

\bibitem{Xue:2012wi}
  W.~Xue, X.~Gao and R.~Brandenberger,
  arXiv:1201.0768 [hep-th].
  W.~Xue, K.~Dasgupta, R.~Brandenberger,
  Phys.\ Rev.\  {\bf D83}, 083520 (2011).
  [arXiv:1103.0285 [hep-th]].

\bibitem{Giddings:2010ui}
  S.~B.~Giddings, M.~S.~Sloth,
  JCAP {\bf 1007}, 015 (2010).
  [arXiv:1005.3287 [hep-th]];\\
  S.~B.~Giddings, M.~S.~Sloth,
  JCAP {\bf 1101}, 023 (2011).
  [arXiv:1005.1056 [hep-th]];\\
  A.~Riotto, M.~S.~Sloth,
  JCAP {\bf 0804}, 030 (2008).
  [arXiv:0801.1845 [hep-ph]].

\bibitem{Akhmedov:2011pj}
  E.~T.~Akhmedov,
  JHEP {\bf 1201}, 066 (2012)
  [arXiv:1110.2257 [hep-th]].

\bibitem{Mottola:1984ar}
  E.~Mottola,
  Phys.\ Rev.\  D {\bf 31}, 754 (1985).

\bibitem{Allen:1985ux}
  B.~Allen,
  Phys.\ Rev.\  D {\bf 32}, 3136 (1985).

\bibitem{Polyakov:2007mm}
  A.~M.~Polyakov,
  Nucl.\ Phys.\ B {\bf 797}, 199 (2008)
  [arXiv:0709.2899 [hep-th]].

\bibitem{Polyakov:2009nq}
  A.~M.~Polyakov,
  Nucl.\ Phys.\ B {\bf 834}, 316 (2010)
  [arXiv:0912.5503 [hep-th]].

\bibitem{Polyakovtalk} A.Polyakov, ``The Dark and the Red'', http://hep.caltech.edu/ym35/

\bibitem{Akhmedov:2008pu}
  E.~T.~Akhmedov and P.~V.~Buividovich,
  Phys.\ Rev.\ D {\bf 78}, 104005 (2008)
  [arXiv:0808.4106 [hep-th]];\\
  E.~T.~Akhmedov, P.~V.~Buividovich and D.~A.~Singleton,
  To appear in Rus. Yad. Fiz.,  arXiv:0905.2742 [gr-qc].

\bibitem{Akhmedov:2009vh}
  E.~T.~Akhmedov and E.~T.~Musaev,
  New J.\ Phys.\  {\bf 11}, 103048 (2009)
  [arXiv:0901.0424 [hep-ph]];\\
  E.~T.~Akhmedov and Ph.~Burda,
  Phys.\ Lett.\ B {\bf 687}, 267 (2010)
  [arXiv:0912.3435 [hep-th]].

\bibitem{Akhmedov:2012hk}
  E.~T.~Akhmedov and A.~V.~Sadofyev,
  arXiv:1201.3471 [hep-th].

\bibitem{Leblond}
  D.~P.~Jatkar, L.~Leblond, A.~Rajaraman,
  [arXiv:1107.3513 [hep-th]].

\bibitem{Kamenev} A.Kamenev, ``Many-body theory of non-equilibrium systems'',  arXiv:cond-mat/0412296.

\bibitem{LL}
L.~D.~Landau and E.~M.~Lifshitz, Vol. 10 (Pergamon Press, Oxford, 1975).

\bibitem{Marolf:2010zp}
  D.~Marolf and I.~A.~Morrison,
  arXiv:1006.0035 [gr-qc];\\
  D.~Marolf, I.~A.~Morrison,
  [arXiv:1104.4343 [gr-qc]];\\
  A.~Higuchi, D.~Marolf, I.~A.~Morrison,
  Phys.\ Rev.\  {\bf D83}, 084029 (2011).
  [arXiv:1012.3415 [gr-qc]];\\
  D.~Marolf, I.~A.~Morrison,
  [arXiv:1010.5327 [gr-qc]].

\bibitem{vanderMeulen:2007ah}
  M.~van der Meulen and J.~Smit,
  JCAP {\bf 0711}, 023 (2007)
  [arXiv:0707.0842 [hep-th]].

\bibitem{Kundu:2011sg}
   S.~Kundu,
   JCAP {\bf 1202}, 005 (2012)
   [arXiv:1110.4688 [astro-ph.CO]].

\end{thebibliography}
\end{document}